\documentclass[preprint]{revtex4-1}
\usepackage{graphicx}
\usepackage{dcolumn}
\usepackage{bm}
\usepackage{amsmath,amssymb}
\usepackage[english]{babel}

\input{tcilatex}

\begin{document}

\title[Extending the ADM formalism to Weyl geometry]{Extending the ADM
formalism to Weyl geometry}
\author{A.B. Barreto}
\email{adrianobraga@fisica.ufpb.br}
\author{T.S. Almeida}
\email{talmeida@fisica.ufpb.br}
\author{C. Romero}
\email{cromero@fisica.ufpb.br}

\begin{abstract}
In order to treat quantum cosmology in the framework of Weyl spacetimes we
take the first step of extending the Arnowitt-Deser-Misner formalism to Weyl
geometry. We then obtain an expression of the curvature tensor in terms of
spatial quantities by splitting spacetime in (3+1)-dimensional form. We next
write the Lagrangian of the gravitation field based in Weyl-type gravity
theory. We extend the general relativistic formalism in such a way that it
can be applied to investigate the quantum cosmology of models whose
spacetimes are endowed with a Weyl geometrical structure.
\end{abstract}

\date{\today}
\maketitle

\affiliation{Departamento de F\'{i}sica, Universidade Federal da
Para\'{i}ba,\\ Jo\~{a}o Pessoa, PB 58059-970, Brazil}

\affiliation{Departamento de F\'{i}sica, Universidade Federal da
Para\'{i}ba,\\ Jo\~{a}o Pessoa, PB 58059-970, Brazil}

\affiliation{Departamento de F\'{i}sica, Universidade Federal da
Para\'{i}ba,\\ Jo\~{a}o Pessoa, PB 58059-970, Brazil}


\section{Introduction}

\begin{quote}
``\textit{Quantum cosmology is the application of quantum theory to the
dynamical systems describing closed cosmology.}'' (J.J. Halliwell)\cite%
{Halliwell}
\end{quote}

It is with the sentence above that we would like to begin this work, because
it appropriately summarizes the approach to quantum cosmology. Another
sentence which we could quote to justify, in part, our interest in the Weyl
geometry follows from Dirac's words:

\begin{quote}
``\textit{It appears as one of the fundamental principles of nature that the
equations expressing the basic laws of physics should be invariant under the
widest possible group of transformations.}'' (P.A.M. Dirac)\cite{Dirac}
\end{quote}

In this work, we consider a gravity theory formulated in the language of
Weyl geometry. Our aim is to develop a formalism that can further be applied
to quantum cosmology. In this geometric structure it is given a purely
geometric scalar field that could play a role to address some issues, such
as the \textit{time problem}\cite{Islam, Kiefer}. Clearly the first step to
carry out this program is to extend the Arnowitt-Deser-Misner formalism (ADM
formalism)\cite{ADM} from Riemannian to Weyl geometry.

In this paper, we briefly review the main points of the Weyl geometry, in
particular those related to the so-called Weyl integrable geometries.
Furthermore, we apply the formalism to some examples in order to find the
role of the Weyl field as a canonical variable of the system. 
\cite{footnote2}

\section{Weyl Geometry}

The geometry conceived by Weyl is a simple generalization of Riemannian
geometry\cite{Weyl, Pauli}. Instead of postulating that the covariant
derivative of the metric tensor $g$ is zero, we assume the more general
condition 
\begin{equation}
\nabla_{\alpha}g_{\mu\nu} = \sigma_{\alpha}g_{\mu\nu},
\label{NonmetricityCondition}
\end{equation}
where $\sigma_{\alpha}$ denotes the components, with respect to a local
coordinates basis $\lbrace \partial_{\alpha} \rbrace$, of a 1-form field $%
\sigma$ defined on the manifold $\mathcal{M}$. This represents a
generalization of the Riemannian condition of compatibility between the
connection $\nabla$ and the metric tensor $g$, namely, the requirement that
the length of a vector remains unaltered by parallel transport. If $\sigma$
vanishes, then (\ref{NonmetricityCondition}) reduces to the familiar
Riemannian metricity condition. It is noteworthy that the Weyl condition (%
\ref{NonmetricityCondition}) remains unchanged when we perform the following
simultaneous transformations in $g$ and $\sigma$: 
\begin{eqnarray}
\bar{g} &=& \text{e}^{f} g,  \label{TransfWeyl1} \\
\bar{\sigma} &=& \sigma + df,  \label{TransfWeyl2}
\end{eqnarray}
where $f$ is an arbitrary scalar function defined on $\mathcal{M}$. These
are known in the literature as Weyl transformations.

If $\sigma = d\phi$, where $\phi$ is a scalar field, then we have what is
called a \textit{Weyl integrable manifold}. In this particular case, the
transformations (\ref{TransfWeyl1}) and (\ref{TransfWeyl2}) become 
\begin{eqnarray}
\bar{g} &=& \text{e}^{f} g,  \label{wistframe1} \\
\bar{\phi} &=& \phi + f.  \label{wistframe2}
\end{eqnarray}
In this paper, we consider only Weyl integrable manifolds.

If the connection $\nabla$ is assumed to be torsionless, then by virtue of
condition (\ref{NonmetricityCondition}) it gets completely determined by $g$
and $\phi$. Indeed, a straightforward computation shows that the components
of the affine connection with respect to an arbitrary vector basis are
completely given by 
\begin{equation}
\Gamma^{\alpha}_{\mu\nu} = \lbrace^{\alpha}_{\mu\nu}\rbrace - \frac{1}{2}
g^{\alpha\beta}\left( g_{\beta\mu}\phi_{\nu} + g_{\beta\nu}\phi_{\mu} -
g_{\mu\nu}\phi_{\beta}\right),  \label{WeylConnection}
\end{equation}
where $\lbrace^{\alpha}_{\mu\nu}\rbrace$ represents the Christoffel symbols
and we are denoting $\phi_{\mu} \doteq \partial_{\mu}\phi$. The affine
connection given in (\ref{WeylConnection}) is called \textit{Weyl connection}
and it is invariant under the Weyl transformations. A nice account of Weyl's
ideas, as well as the refutation of his gravitational theory, may be found
in reference\cite{CRomero}. 

\section{The splitting of the spacetime}

As is well known, the first obstacle that we have to deal with in the
quantum cosmology approach is the incompatibility between geometrical
gravity theories and the Hamiltonian or canonical formalism: geometrical
gravity theories treat space and time on the same footing, while the
Hamiltonian formalism is given in terms of a set of canonical variables
taking value at a given instant of time. Therefore, we need to adapt the
description of the spacetime dynamics given by the geometrical theory of
gravity to the Hamiltonian approach\cite{Gambini}. This canonical
formulation requires a spacetime splitting\cite{Bojowald}. The line element,
which carries out the splitting of the (3+1)-spacetime into foliated $3$%
-dimensional space-like hypersurfaces, is well known in the literature\cite%
{ADM, Pad, Bojowald, Halliwell} and is written as 
\begin{equation}
ds^{2} = (N^{i} N_{i} - N^{2})dt^{2} + 2 N_{j}dx^{j}dt + h_{ij}dx^{i}dx^{j},
\label{4metric}
\end{equation}
where the scalar function $N$ is called the \textit{lapse function}, $N^{i}$
refers to the components of a spatial 3-vector, called \textit{shift vector}%
, and $h_{ij}$ is the induced 3-metric on $\Sigma_{t}$. It is simple to
identify the metric components in terms of $N, N^{i}$ and $h_{ij}$: 
\begin{equation}
g_{00} = N^{i} N_{i} - N^2 \;\;\;\; , \;\;\;\; g_{0j}=N_j \;\;\;\; ,
\;\;\;\; g_{ij}=h_{ij},  \label{MetricComp}
\end{equation}
with the contravariant components 
\begin{equation}
g^{00} = - \frac{1}{N^{2}} \;\;\; , \;\;\; g^{0j} = \frac{N^j}{N^2} \;\;\; ,
\;\;\; g^{ij} = h^{ij} - \frac{N^{i} N^{j}}{N^2}
\end{equation}
and $\sqrt{-g} = N \sqrt{h}$. Therefore, the splitting of the spacetime is
achieved mathematically by foliating the spacetime in a series of space-like
hypersurfaces $\Sigma_{t}$, labeled by a coordinate $t$ through a function $%
t(x^\mu)$ defined on the spacetime\cite{Pad}. The induced metric on $%
\Sigma_{t}$ can be express in four-dimensional notation by  
\begin{equation}
h_{\mu\nu} = g_{\mu\nu} + n_{\mu} n_{\nu}.  \label{IndMetric}
\end{equation}
Taking into account that $n_{\mu} \equiv (-N, \vec{0})$, it is
straightforward to obtain $h_{00} = N^{i} N_{i}$, $h_{0i} = N_{i}$ and $%
h_{ij} = g_{ij}$. Although $h_{\mu\nu}$ is a space-like $3$-metric, $h_{00}$
and $h_{0i}$ are non-null because $g_{0i} \neq 0$. These non-null components
are required to cancel the effects of $g_{0i}$. Furthermore, by definition
of the induced metric (\ref{IndMetric}), we have $h^{\mu}_{\; \nu}n^{\nu} = 0
$ as an orthogonality relation.

Besides, taking into consideration the non-metricity Weylian condition (\ref%
{NonmetricityCondition}) and the normalization relation $n_{\mu}n^{\mu} = -1$%
, we can obtain 
\begin{equation}
n^{\mu} \nabla_{\nu} n_{\mu} = - \frac{1}{2} \phi_{\nu} \;\;\;\; \text{and}
\;\;\;\; n_{\mu} \nabla_{\nu} n^{\mu} = \frac{1}{2}\phi_{\nu},
\label{NonMetricCond}
\end{equation}
The relations above, result from the difference between Riemannian and
Weylian geometries, and are essential to turn the geometric quantities
associated to the gravitational field into functions that depend on the Weyl
field. Naturally, the known results of general relativity will arise in the
limit when the Weyl field vanishes.


\section{Extrinsic Curvature and Gauss-Codazzi equations}

The induced metric $h_{\mu\nu}$ itself is an intrinsic quantity, and as a
metric it allows us to define a unique covariant derivative operator $D_{\mu}
$ on $\Sigma_{t}$, by using the projection tensor $h^{\alpha}_{\;\beta} =
\delta^{\alpha}_{\;\beta} + n^{\alpha}n_{\beta}$ on the $4$-dimensional
covariant derivative $\nabla_{\mu}X_{\nu}$: 
\begin{equation}
D_{\alpha}X_{\beta} \doteq
h^{\mu}_{\;\alpha}h^{\nu}_{\;\beta}\nabla_{\mu}X_{\nu}.  \label{SpatialD}
\end{equation}
Hence, assuming the rule $D_{\alpha}f = h^{\beta}_{\;\alpha}\nabla_{\alpha}f$%
, where $f$ is a scalar function, we can determine the action of $D_{\mu}$
on the contravariants vectors, 
\begin{equation}
D_{\alpha}V^{\beta} = h^{\mu}_{\;\alpha}h^{\beta}_{\;\nu}\nabla_{\mu}V^{\nu}.
\label{SpatialD2}
\end{equation}

It is important to note that the spatial covariant derivative of the induced
metric also reflects the consequences of the non-metricity condition.
Indeed, it is not difficult to see that 
\begin{equation}
D_{\alpha}h_{\mu\nu} = h^{\beta}_{\;\alpha}\phi_{\beta} h_{\mu\nu},
\label{nonmetricS}
\end{equation}
since $h^{\lambda}_{\;\nu}n_{\lambda} = 0$ and $\nabla_{\beta}g_{\gamma%
\lambda} = \phi_{\beta}g_{\gamma\lambda}$. Thus, if $h^{\beta}_{\;\alpha}%
\phi_{\beta}$ is the gradient of the Weyl field projected on $\Sigma_{t}$,
we have the interesting case where the metricity condition holds on $%
\Sigma_{t}$. This case occurs when the Weyl field depends only on the time
coordinate, i.e., $\phi_{\mu} \equiv \phi_{\mu}(t)$, which is a plausible
hypothesis for homogeneous and isotropic cosmological models. Therefore,
since $\phi_{\mu} \propto n_{\alpha}$ we have $h^{\beta}_{\;\alpha}\phi_{%
\beta} = 0$, and (\ref{nonmetricS}) becomes 
\begin{equation}
D_{\alpha}h_{\mu\nu} = 0.  \label{metricityS}
\end{equation}

Now we are able to describe the intrinsic features of the spatial slices,
since we have the induced metric and a covariant derivative intrinsic to $%
\Sigma_{t}$. However, to obtain the complete information about the structure
of the spacetime, we need to know how the hypersurfaces are embedded in the
spacetime. Intuitively we expect that this information is contained in the
manner how $\Sigma_{t}$ vary from point to point. This variation is given by
the so-called \textit{extrinsic curvature} of $\Sigma_{t}$, defined by 
\begin{equation}
\mathcal{K}_{\alpha\beta} \doteq - h^{\mu}_{\alpha} h^{\nu}_{\beta}
\nabla_{\mu} n_{\nu}.  \label{extrcurvW}
\end{equation}

In Riemannian terms, we have the useful expression 
\begin{equation}
\mathcal{K}_{\alpha\beta} = \widetilde{K}_{\alpha\beta} + \frac{1}{2}
h_{\alpha\beta}\xi,  \label{extrcurvW2}
\end{equation}
where $\xi \doteq n_\mu \phi^\mu$ is the normal component of the Weyl field
and $\widetilde{K}_{\alpha\beta} \doteq -h^{\mu}_{\alpha} h^{\nu}_{\beta}%
\widetilde{\nabla}_{\mu} n_{\nu}$ is the Riemannian extrinsic curvature. We
are denoting the quantities built with the Riemannian connection with a
tilde ($\; \widetilde{}\; $). For instance, $\widetilde{\nabla}%
_{\mu}V^{\alpha} \doteq \partial_{\mu}V^{\alpha} +
\lbrace^{\alpha}_{\mu\nu}\rbrace V^{\nu}$ and $\widetilde{\Box} \doteq
g^{\mu\nu}\widetilde{\nabla}_{\mu}\widetilde{\nabla}_{\nu}$, where the
latter denotes the Laplace-Beltrami operator.

In other words, the Weylian extrinsic curvature differs from the riemannian
one by the projections of the Weyl field (its gradient in fact) in the
normal direction. The extrinsic curvature $\mathcal{K}_{\mu\nu}$ is a
symmetric tensor\cite{footnote1}, $\mathcal{K}_{\mu\nu} = \mathcal{K}%
_{\nu\mu}$, and its contravariant components are purely spatial. From
equations (\ref{IndMetric}) and (\ref{extrcurvW}), and defining the $4$%
-vector acceleration as 
\begin{equation}
a_{\mu} = n^{\alpha}\nabla_{\alpha}n_{\mu},  \label{4-Acceleration}
\end{equation}
we conclude that the $4$-dimensional covariant derivative of the normal
vector can be decomposed in the following terms 
\begin{equation}
-\nabla_{\mu}n_{\nu} = \mathcal{K}_{\mu\nu} + n_{\mu}a_{\nu} - \frac{1}{2}%
n_{\nu}\phi_{\alpha}h^{\alpha}_{\; \mu}.  \label{KDecomposed}
\end{equation}

This equation is helpful to write the curvature tensor in terms of $\mathcal{%
K}_{\mu\nu}$. Moreover, it is useful to obtain the expression of the spatial
components of $\mathcal{K}_{\nu\mu}$ in terms of $\lbrace N, N^{i}, h_{ij},
\phi_{\mu}\rbrace$. Thus, by definition, we have 
\begin{equation}
\mathcal{K}_{ij} = \frac{1}{N}\left[D_{(i}N_{j)} - \phi_{(i}N_{j)} + \frac{1%
}{2}\left( h_{ij}\dot{\phi} - \dot{h}_{ij}\right) \right],  \label{ExtCurvS}
\end{equation}
where parentheses are being used to denote symmetrization of indices 
\begin{equation}
S_{(\mu\nu)} \equiv \frac{1}{2} \left( S_{\mu\nu} + S_{\nu\mu} \right),
\label{SymmNotation}
\end{equation}
and dots represent the time coordinate derivatives, i.e., $\dot{\phi} \equiv
\partial_{0}\phi$ and $\dot{h}_{ij} \equiv \partial_{0} h_{ij}$. Thus, we
note the presence of a time derivative of the Weyl field in the expression
of the extrinsic curvature, which implies that the Weyl field plays the role
of a canonical variable of the system. Besides, it is clear that in a
particular coordinate system in which $N^{i}$ is null, we would get 
\begin{equation}
\mathcal{K}_{ij} = \frac{1}{2N}\left(h_{ij}\dot{\phi} - \dot{h}_{ij}\right).
\label{ExtCurvS2}
\end{equation}

Once a particular foliation of the Weyl spacetime is given, $h_{\mu\nu}$ and 
$\mathcal{K}_{ij}$ contain the information about the intrinsic and extrinsic
features of the $\Sigma_{t}$. Thus, in order to write the $4$-dimensional
curvature tensor $R^{\delta}_{\;\rho\sigma\gamma}$ in terms of $\mathcal{K}%
_{\mu\nu}$ and the curvature tensor of the $3$-dimensional hypersurface $%
^{(3)} R^{\alpha}_{\;\beta\mu\nu}$, we must use the \textit{Gauss-Codazzi
equations}: 
\begin{equation}
^{(3)}R_{\alpha\beta\mu\nu} =
h^{\delta}_{\;\alpha}h^{\rho}_{\;\beta}h^{\sigma}_{\;\mu}h^{\gamma}_{\;%
\nu}R_{\delta\rho\sigma\gamma} + \mathcal{K}_{\alpha\mu}\mathcal{K}%
_{\beta\nu} - \mathcal{K}_{\alpha\nu}\mathcal{K}_{\beta\mu}.  \label{GCE}
\end{equation}

Finally, the above expression (\ref{GCE}) allows us to deduce the ADM
Lagrangian corresponding to a given gravity theory. In this paper, we will
be concerned with theories whose Lagrangians consists of the curvature
scalar $R$ and a kinetic term with respect to the Weyl field. 

\section{The ADM Lagrangian in Weyl-type theories}

Until now we have obtained the appropriate generalizations of some geometric
quantities from a Riemannian context to the case of Weyl integrable
manifold. This enables us to deduce the ADM Lagrangian for any Weyl-type
gravity theory. Examples of the latter are a proposal known in the
literature as the \textit{Weyl integrable spacetime} gravity (WIST)\cite%
{Novello, WIST}, described by the action 
\begin{equation}
\mathcal{S}_{wist} = \int \sqrt{-g}\left( R + \omega
\phi_{\alpha}\phi^{\alpha}\right)d^{4}x,  \label{wist}
\end{equation}
and also a geometric approach to Brans-Dicke theory (GBD)\cite{GBD}, which
postulates the action 
\begin{equation}
\mathcal{S}_{GBD} = \int \sqrt{-g}\text{e}^{-\phi}\left( R + \omega
\phi_{\alpha}\phi^{\alpha}\right)d^{4}x,  \label{gBD}
\end{equation}
where $R$ is the Weyl curvature scalar, $\phi$ is the Weyl field and $\omega$
is dimensionless parameter. In order to obtain $R$ in terms of the extrinsic
and intrinsic curvatures we contract $R_{\lambda\nu\sigma\tau}$ with $%
h_{\alpha\beta}$, which gives 
\begin{equation}
R = h^{\lambda\sigma}h^{\nu\tau}R_{\lambda\nu\sigma\tau} -
2n^{\lambda}n^{\sigma}R_{\lambda\sigma}.
\end{equation}

It can be shown that 
\begin{equation}
R = \; ^{(3)}R - \mathcal{K}^{\gamma}_{\;\nu}\mathcal{K}^{\nu}_{\;\gamma} + 
\mathcal{K}^{2} + 2\left(\nabla_{\mu} + \phi_{\mu}\right)A^{\mu},
\label{RDecompWeyl}
\end{equation}
where we define $A^{\mu} \doteq a^{\mu}+ n^{\mu}(\mathcal{K}-\frac{1}{2}\xi)$%
, $\mathcal{K}$ is the trace of $\mathcal{K}_{\mu\nu}$ and $^{(3)}R$
represents the 3-dimensional curvature scalar, i.e., the curvature scalar
calculated with the 3-metric $h_{ij}$. On the other hand, we have 
\begin{equation}
\nabla_{\alpha}A^{\alpha} = \widetilde{\nabla}_{\alpha} A^{\alpha} - \frac{1%
}{2}\left(\delta^{\alpha}_{\;\alpha}\phi_{\beta} +
\delta^{\alpha}_{\;\beta}\phi_{\alpha} -
g_{\alpha\beta}\phi^{\alpha}\right)A^{\beta},  \label{CovDerSplitting}
\end{equation}
which leads to 
\begin{equation}
\widetilde{\nabla}_{\alpha} A^{\alpha} \equiv
\left(\nabla_{\alpha}+2\phi_{\alpha}\right)A^{\alpha}.  \label{DiverRiemn}
\end{equation}
In this way, we can write 
\begin{equation}
R = L_{ADMW} + 2\left(\nabla_{\mu} + 2\phi_{\mu}\right)A^{\mu}
\label{R-Decomp-ADM}
\end{equation}
where the latter term in (\ref{R-Decomp-ADM}) is the \textit{Riemannian
divergence} shown in (\ref{DiverRiemn}). Finally the ADM Lagrangian in Weyl
integrable spacetime will be given by 
\begin{equation}
L_{ADMW} \doteq \; ^{(3)}R - \mathcal{K}^{\gamma}_{\;\nu}\mathcal{K}%
^{\nu}_{\;\gamma} + \mathcal{K}^{2} - 2\phi_{\mu}A^{\mu}.  \label{RWADM}
\end{equation}

Another way to obtain an expression of the ADM Lagrangian in a Weyl
integrable spacetime is by decomposing $R$ in terms of Riemannian quantities
and $\phi$.\cite{CRomero}. Indeed, from 
\begin{equation}
R_{\mu\nu} \equiv \widetilde{R}_{\mu\nu} - \frac{3}{2}\widetilde{\nabla}%
_{\nu}\phi_{\mu} + \frac{1}{2}\widetilde{\nabla}_{\mu}\phi_{\nu} - \frac{1}{2%
}\phi_{\mu}\phi_{\nu} + \frac{1}{2}g_{\mu\nu}\left(
\phi^{\alpha}\phi_{\alpha}-\widetilde{\nabla}_{\alpha}\phi^{\alpha}\right),
\label{eqRmunuw-r}
\end{equation}
we obtain, 
\begin{equation}
R = \widetilde{R} -3\widetilde{\Box}\phi +\frac{3}{2}g^{\mu\nu}\phi_{\mu}%
\phi_{\nu}.  \label{eqRw-r}
\end{equation}

Therefore, given an arbitrary Weyl-like action 
\begin{equation}
\mathcal{S}_{G} = \int \sqrt{-g} f(\phi) \left[ R +
j(\phi)g^{\mu\nu}\phi_{\mu}\phi_{\nu}\right]d^{4}x ,  \label{S_G}
\end{equation}
where $f(\phi)$ and $j(\phi)$ are functions that ``label'' a specific
theory, we can write (\ref{S_G}) in the form 
\begin{equation}
\mathcal{S}_G = \int \sqrt{-g}f(\phi)\left[\widetilde{\mathcal{L}}_{ADMR} +
\lambda(\phi) g^{\mu\nu}\phi_{\mu}\phi_{\nu}\right]d^{4}x - 2 \int \sqrt{-g}%
f(\phi)\left[\widetilde{\Box}\phi + \widetilde{\nabla}_{\alpha}\left( \xi
n^{\alpha}\right)\right]d^{4}x ,  \label{S-ADM+TD}
\end{equation}
where we have redefined the parameter to $\lambda(\phi) \doteq \frac{3}{2}+
j(\phi)$, and used that\cite{Bojowald}, 
\begin{equation}
\widetilde{R} \equiv \widetilde{\mathcal{L}}_{ADMR} + 2 \widetilde{\nabla}%
_{\mu}\left( \widetilde{a}^{\mu} + n^{\mu}\widetilde{K} \right) .
\label{R-Riemann-ADM}
\end{equation}
The Riemannian ADM Lagrangian $\widetilde{\mathcal{L}}_{ADMR}$ is well known
in the literature\cite{Pad, Bojowald, Halliwell}, and is given by 
\begin{equation}
\widetilde{\mathcal{L}}_{ADMR} \doteq \; ^{(3)}\widetilde{R} - \widetilde{K}%
^{\mu}_{\;\; \nu}\widetilde{K}^{\nu}_{\;\; \mu} + \widetilde{K}^{2}.
\label{L_ADMR}
\end{equation}
For simplicity, we group the terms with divergences in the definition of the
function $D_{iv}$: 
\begin{equation}
D_{iv} \doteq \int \sqrt{-g}f(\phi)\left[\widetilde{\Box}\phi + \widetilde{%
\nabla}_{\alpha}\left( \xi n^{\alpha}\right)\right]d^{4}x ,  \label{TD}
\end{equation}
where in this (3+1) foliation, we have 
\begin{equation}
\xi \doteq g^{\mu\nu}n_{\mu}\phi_{\nu} \equiv \frac{\dot{\phi}}{N^2}.
\end{equation}
We can work out the surface terms that appear in (\ref{TD}) and thus write 
\begin{equation}
\mathcal{S}_G = \int \sqrt{-g}f(\phi)\left[\widetilde{\mathcal{L}}_{ADMR} +
\lambda\phi_{\mu}\phi^{\mu}\right]d^{4}x + 2 F_{\phi} + \text{\textit{%
Surface Terms}},  \label{S_Final}
\end{equation}
with 
\begin{equation}
F_{\phi}\doteq \int \sqrt{-g}\left[f^{\prime }\left( \frac{\dot{\phi}}{N^2}
+ \phi\widetilde{\Box}\phi\right) + \phi f^{\prime \prime
}\phi_{\mu}\phi^{\mu}\right]d^{4}x ,  \label{F_phi}
\end{equation}
and $f^{\prime }\doteq \frac{\partial f}{\partial \phi}$, $f^{\prime \prime
}\doteq \frac{\partial^2 f}{\partial \phi^2}$.

In this way we have shown that the Weyl field is identified with a canonical
variable, since there is canonical momentum conjugated to it. The functional
form of this momentum depends on $f(\phi)$ and $j(\phi)$. In the case of
WIST, where $f(\phi) = \text{const.}$ and $\lambda = \omega$, we get 
\begin{equation}
\mathcal{S}_{G} = \int \sqrt{-g} \left( \widetilde{\mathcal{L}}_{ADMR} +
\omega g^{\mu\nu}\partial_{\mu}\phi\partial_{\nu}\phi\right)d^{4}x .
\label{S-ADMWIST}
\end{equation}
Let us note that the expression for the action (\ref{S-ADMWIST}) is
identical (with $\omega = -1/2$) to the case of quantum cosmology in the
context of general relativity minimally coupled to a massless scalar field
in a FLRW cosmological models\cite{Lemos}. We are now ready to proceed to
the quantization of FLRW cosmological models determined by the Weyl
integrable spacetime gravity theory. We leave this for future work. 
\begin{acknowledgments}
The authors would like to thank CAPES/CNPq and the organizing committee of II-Cosmosur for financial support.  We are grateful to Dr M.L. Pucheu for helpful discussions and suggestions.
\end{acknowledgments}


\nocite{*}

\begin{thebibliography}{99}
\bibitem{Halliwell} J. J. Halliwell, "Introductory lectures on quantum
cosmology" (1990), arXiv:0909.2566 [gr-qc].

\bibitem{Dirac} P. A. M. Dirac, "Long Range Forces and Broken Symmetries",
Proc.Roy.Soc.Lond. \textbf{A333} 403-418 (1973).

\bibitem{Islam} C. J. Islam, \textit{Integrable Systems, Quantum Groups, and
Quantum Field Theories} (Springer Netherlands, 1993).

\bibitem{Kiefer} C. Kiefer, "Conceptual issues in quantum cosmology",
Lect.Notes Phys. \textbf{541} 158-187 (2000), arXiv:gr-qc/9906100 [gr-qc].

\bibitem{ADM} R. L. Arnowitt and S. Deser and C. W. Misner, "The Dynamics of
general relativity", Gen.Rel.Grav. \textbf{40} 1997-2027 (2008),
arXiv:gr-qc/0405109 [gr-qc].

\bibitem{footnote2} We are going to work with the notation where the
spacetime coordinates are labeled by Greek indices and the 3-space
coordinates are labeled by Latin indices, so $x^{\mu} \equiv (x^{0}, x^{i})$%
, such that $x^{0}$ is the time coordinate. Furthermore, we adopt the
following convention in the definition of the curvature tensor $%
R^{\alpha}_{\; \beta \mu \nu} \doteq \partial_{\nu}\Gamma^{\alpha}_{\; \beta
\mu} - \partial_{\mu}\Gamma^{\alpha}_{\; \beta \nu} + \Gamma^{\lambda}_{\;
\beta \mu}\Gamma^{\alpha}_{\; \lambda \nu} - \Gamma^{\lambda}_{\; \beta
\nu}\Gamma^{\alpha}_{\; \lambda \mu}$. In order to have a positive-definite
induced metric over the space-like hypersurface $h_{ij}$, we take the
signature diag$(g) = (-, +, +, +)$.

\bibitem{Weyl} H. Weyl, \textit{Space, Time, Matter} (Dover, 1952).

\bibitem{Pauli} W. Pauli, \textit{Theory of Relativity} (Dover, 1958).

\bibitem{CRomero} C. Romero and J. B. Fonseca-Neto and M. L. Pucheu,
"General Relativity and Weyl Geometry", Class.Quant.Grav. \textbf{29},
155015 (2012), arXiv:1201.1469 [gr-qc].

\bibitem{Gambini} R. Gambini and J. Pullin, \textit{A First Course in Loop
Quantum Gravity} (OUP Oxford, 2011).

\bibitem{Bojowald} M. Bojowald, \textit{Canonical Gravity and Applications}
(Cambridge University Press, 2011).

\bibitem{Pad} T. Padmanabhan, \textit{Gravitation: Foundations and Frontiers}
(Cambridge University Press, 2010).

\bibitem{footnote1} To show that $\mathcal{K}_{\mu\nu}$ is a symmetric
tensor, one must use the Frobenius' theorem.

\bibitem{Novello} M. Novello and L. A. R. Oliveira and J. M. Salim and E.
Elbaz, "Geometrized Instantons and the Creation of the Universe",
Int.J.Mod.Phys. \textbf{D1} 641-677 (1992).

\bibitem{WIST} J. M. Salim and S. L. Sautu, "Gravitational theory in Weyl
integrable spacetime", Class.Quant.Grav. \textbf{13} 353-360 (1996).

\bibitem{GBD} T. S. Almeida and M. L. Pucheu and C. Romero and J. B.
Formiga, "From Brans-Dicke gravity to a geometrical scalar-tensor theory",
Phys.Rev. \textbf{D89} 064047 (2014), arXiv:1311.5459 [gr-qc].

\bibitem{Lemos} E. M. Barboza Jr. and N. A. Lemos, "Can quantum
gravitational effects influence the entire history of the Universe?",
Phys.Rev. \textbf{D78} 023504 (2008), arXiv:0805.3971 [gr-qc].

\end{thebibliography}

\end{document}